\documentclass[12 pt]{article}

\usepackage{graphicx}
\usepackage{rotating} 
\usepackage{fullpage,amsmath,amssymb,latexsym}
\usepackage{multirow}
\usepackage{natbib}

\begin{document}

\title {Measuring Jupiter's water abundance by Juno: \\ the link between interior and formation models}
\author
  {Ravit Helled$^1$ and
   Jonathan Lunine$^2$\\
      $^1$Department of Geophysics and Planetary Sciences, 
Tel-Aviv University, Israel.\\
$^2$ Department of Astronomy,  
Cornell University, Ithaca, NY, USA.}

\date{}
\maketitle
\begin{abstract}
The {\it Juno} mission to Jupiter is planned to measure the water abundance in Jupiter's atmosphere below the cloud layer. 
This measurement is important because it can be used to reveal valuable information on Jupiter's origin and its composition. In this paper we discuss the importance of this measurement, the challenges in its interpretation, and address how it can be connected to interior and formation models of Jupiter.
\end{abstract}


\section{Introduction}

Jupiter is composed mostly of hydrogen and helium and a smaller fraction of {\it heavy elements}. These heavy elements are important because they are the key for understanding Jupiter's origin and giant planet formation in general (e.g., Pollack et al. 1996; D'Angelo et al., 2011). The exact mass of these heavy elements and their distribution within Jupiter, however, are not well constrained. The NASA {\it Juno} mission to Jupiter is designed to improve our understanding of the planet's internal structure, and possibly, its formation history (Bolton, 2006). Among other measurements, {\it Juno} will determine the water abundance in Jupiter's gaseous envelope. Jupiter's water abundance will be measured using the Microwave Radiometer which will probe Jupiter's deep atmosphere ($\sim$ 100 bar) at radio wavelengths  ranging from 1.3 to 50 cm using six separate radiometers to measure the thermal emissions. This measurement can be used to determine the water abundance in Jupiter's convective zone (and therefore, its oxygen-to-hydrogen ratio, hereafter, $(O/H)_{J}$). \footnote{This measurement can be used to determine the {\it global} $(O/H)_{J}$ only if Jupiter is indeed adiabatic and homogeneously mixed. In addition, it should be noted that what is measured is the water abundance and not the $(O/H)_{J}$ directly.} 

In addition, {\it Juno} will determine Jupiter's gravitational field (gravitational moments) to a very high accuracy therefore providing tighter constraints on its density distribution, and hence, its composition.
Accurate measurements of Jupiter's water abundance below the cloud-level combined with measurements of its gravitational field can be used to better constrain both interior and formation models of Jupiter but their interpretation should also be treated with great caution as discussed below. 
\par

\section{Formation Models}

\subsection{Formation by Core Accretion}
Core accretion, the standard model for giant planet formation, suggests that giant planets are formed by planetesimal accretion and core formation followed by accretion of a gaseous envelope (e.g., Pollack et al. 1996).  
Formation models of Jupiter must be able to form a Jupiter-mass planet within the right timescale, i.e., before the gas disk dissipates ($< 10^7$ years), and within reasonable assumptions regarding its formation location and the physical/chemical properties in the solar nebula in which the planet is formed. In addition, the composition of the forming planet should be consistent with the composition derived from interior models that use the observed physical properties of Jupiter today. Therefore, the composition inferred from formation models of Jupiter should be consistent with the heavy element mass derived from structure models.  \par
 
Until recently, the core accretion model has demonstrated the formation of giant planets successfully, in the sense that runaway gas accretion was reached within the required timescale, but the models were typically stopped when runaway gas accretion was reached, and mechanisms for terminating the gas accretion, which essentially determines the final mass of the forming planet were not included (Pollack et al., 1996; Alibert et al., 2005a). Updated formation models now account for disk dissipation and/or gap opening, which are mechanisms that can terminate the gas accretion, and core accretion models can now simulate the formation of giant planets with various final masses and compositions. 
In addition to the termination of gas accretion, in order to get a final mass of 1 M$_J$ other model parameters and the physical mechanisms that are included in the model must be adjusted (Movshovitz et al., 2010; Mordasini et al., 2009; 2012).  
Critical parameters that influence the planet's growth are the solid-surface density, the location in which the planet is assumed to form and whether it changes with time (due to migration), and the physical properties of the planetesimals that are accreted by the protoplanet. The solid-surface density is important since the higher it is the faster the planet forms, and in addition the more metal-rich the planet can become. 
Planetesimals are usually assumed to be made of ices (or ice mixed with other materials) since the planet is formed beyond the snow line, where ice condenses, and are often assumed to be $\sim$30\% water ice by mass, with the rest being divided between rocks and `organics'. The planetesimals' sizes are unknown and models consider sizes ranging from a few hundred meters to 100 kilometer.  Small planetesimals and/or planetesimals with low material strengths and/or low density are easier to capture, and therefore, affect the formation timescale of the planets as well as its final composition.  The planetesimals' dynamical properties (velocity, inclination, eccentricity) are also of great importance as they also affect the capture efficiency. 
Nevertheless, despite the uncertainties in planet formation simulations, the requirement to reach a Jupiter-mass within a few million years provides constraints on the possible range of heavy elements within the planet, and also on the properties of the planetesimals accreted by proto-Jupiter. 
\par 

Alibert et al. (2005b) presented a formation model for Jupiter in which migration, disk evolution, and ejection of planetesimals were included. Under the assumption that Jupiter's embryo was formed between 9.2 and 13.5 AU they derived a core mass of $\sim$6 M$_{\oplus}$ and heavy element mass of $\sim$28 M$_{\oplus}$ within the envelope, i.e., a total mass of heavy elements of $\sim$34 M$_{\oplus}$. Lissauer et al. (2009) also presented a formation model for Jupiter which included disk evolution and gap formation but the planet was assumed to form {\it in situ}, i.e. at 5.2 AU. The derived total mass of heavy elements was found to be only $\sim$ 20 M$_{\oplus}$, about 40\% smaller than the one inferred by Alibert et al. (2005b). Besides the different model assumptions mentioned above, the model of Alibert et al. (2005b) includes planetesimal accretion during the gas accretion phase, which leads to a larger enrichment, while in the model of Lissauer et al. (2009) no planetesimals are accreted during that stage. \par

\subsection{Estimating The Water Abundance}
Even the limited comparison we presented above is sufficient to conclude that {\it there is no one unique model of Jupiter's formation even within the framework of the core accretion scenario}, and the final composition of the planet can vary significantly due to the model assumptions. Once the constraints on the heavy-element mass in Jupiter become tighter, which should be possible with {\it Juno}'s measurements, formation models of Jupiter can be better constrained. For example, while the composition of the heavy elements (planetesimals) is unknown, one can derive a rough upper limit on the global water abundance in Jupiter from the available formation models simply by assuming that all the heavy elements in Jupiter are represented by H$_2$O. 
Assuming that the heavy element fraction in Jupiter (by mass) is $f_Z$ and the rest is hydrogen and helium, given by $f_{XY}$ with a solar ratio, we can compute the upper limit of the ratio between the number of oxygen atoms $N_O$ to the number of hydrogen atoms $N_H$, hereafter, $(O/H)_{J}$ by,

\begin{equation} 
O/H= \frac{f_Z/18}{2 f_Z/18 + f_{XY}/(1+4\times10^{-1.01})},
\end{equation}

where, following Lodders (2003) $N_{He}=N_H\times10^{-1.01}$. While Alibert et al. (2005b) predict $(O/H)_{J}/(O/H)_{\odot} = 17.6$ using the Lissauer et al. (2009) model results in $(O/H)_{J}/(O/H)_{\odot}= 8.9$, significantly smaller. This rough estimate, however, provides only the upper bound, since all the heavy elements are represented by water while in reality they should also consist of silicates and organic material. In addition, further complexity arises from the fact that the nature of the ices in planetesimals, and different types of ices (i.e., amorphous vs. clathrate) can lead to very different water enrichments (e.g., Owen et al., 1999, Lunine et al., 2004). \par 

We next consider a more realistic case in which the water is assumed to consist of only 30\% of the total mass of the heavy elements. In that case the $(O/H)_{J}/(O/H)_{\odot} $ is found to be $\sim 4.5$ and sub-solar for the models of Lissauer et al. (2009) and Alibert et al. (2005b), respectively. 
These values are rather different from an enrichment of $\sim 4$ times solar, as measured by the {\it Galileo} probe for the other components in Jupiter's atmosphere such as carbon and nitrogen (e.g., Atreya et al., 2003), and clearly significantly larger than the low value measured by {\it Galileo}, $(O/H)_{J}/(O/H)_{\odot} $=0.25, although the fact that the probe entered a hot spot, a region that is expected to be relatively dry (e.g. Showman and Ingersoll, 1998) makes the measured water abundance simply a lower limit. Despite the remaining uncertainty in the planetesimals' composition it is clear that an accurate determination of $(O/H)_{J}$ can be used to constrain Jupiter's formation models, and the physical conditions of the environment in which Jupiter has formed.

The relation between planetary metallicity and heavy element mass is Jupiter is shown in Figure 1. The top panel simply shows Jupiter's metallicity (Z$_J$) divided by the solar metallicity (Z$_{\odot}$) which is set to be 0.0142 as a function of total heavy element mass. The bottom panel of the figure presents the expected $(O/H)_J/(O/H)_{\odot}$ as a function of (total) heavy element mass (in Earth masses) for four different fractions of water. We consider an extreme case in which the heavy elements consist of pure water (i.e., 100\%, solid curve), and other cases in which the planetesimal composition is 50\% (dashed curve), 30\%(dotted curve) and 15\%(dashed-dotted curve) of water. 
While 30\% might represent the standard cases, there is no reason to exclude low water abundance for Jupiter (see section 4). 
Clearly, a combination of a determination of the total mass of heavy elements in Jupiter {\it together} with an accurate determination of $(O/H)_{J}$, information on the planetesimal composition can be inferred. 
Alternatively, information on the water abundance of the accreted planetesimal could be used to predict the enrichment of Jupiter with water.   \par

While formation models provide estimates for the total mass of heavy elements within the planet, there is an uncertainty regarding the fate, and therefore the distribution, of the accreted heavy elements.  While in the early stages, the planetesimals build up the core, as the planetary embryo accretes a gaseous atmosphere, the planetesimals do not go all the way to the center, but instead, are dissolved in the gaseous envelope. The deposition of heavy elements in the atmosphere depends on the sizes of the planetesimals, and on their compositions. Water is predicted to stay in the atmosphere (Iaroslavitz \& Podolak, 2007), while the refractory materials tend to settle towards the center. Of course, convective mixing can also change the distribution of the heavy elements.  
In order to put limits on the energy released by the planetesimals settling Pollack et al. (1996) considered two extreme cases; in the first the planetesimals are deposited in the upper atmosphere while in the second they reach all the way to the center. However, for the computation of the rate of core growth, it was assumed that all of the high-Z material eventually ended up in the core and that the envelope remained in solar composition. 
This assumption is not strictly correct because it is likely that large portions of the accreted planetesimals are dissolved in the gaseous envelope. This aspect in formation models is very important because the fate of the accreted material determines the atmospheric metallicity, and therefore, the predicted heavy-element distribution. 

In fact, it is found that once the core reaches a mass of $\sim 2 M_{\oplus}$, the bulk of the accreted planetesimals remains in the atmosphere (Hubickyj, O. and Bodenheimer, P., private communication). As a result, we consider two different cases; in case (1) we use the core mass as derived by the models, in which the heavy elements are assumed to join the core, while in case (2) we set the core mass to be 2 $M_{\oplus}$ and assume that the remaining accreted planetesimals stay in the envelope, a scenario that naturally leads to a larger atmospheric enrichment. The resulting $(O/H)_J$ for various formation models are listed in Table 1. 
\par

An alternative to the upper-bounds defined above can be an estimate that accounts for the constraints from atmospheric measurements which were derived from the in situ measurements of the Galileo probe (Atreya et al., 2003). Since formation models provide the total heavy element mass in Jupiter, we can compute the metallicity of its envelope $Z_{J}$ which is related to the elemental abundances by (Nettelmann et al. 2012),  

\begin{equation} 
Z_{J}= \frac{\sum_i \mu_i(N_i/H)}{\mu_H+\mu_{He}(He/H)+\sum_i \mu_i(N_i/H)},
\end{equation}

where $\mu_i$ is the atomic weight of species $i$ and $N_i/H$ are the measured abundances in Jupiter's atmosphere.  
The elemental abundances measured in the tropospheres of Jupiter and their comparison to solar are taken from Atreya et al. (2003) and Lodders (2003). For this estimate the heavy elements in the envelope are assumed to be homogeneously mixed. The helium is assumed to be in solar ratio and {\it not} depleted and the neon measurement is also not included since, like helium, the measured depletion is likely to be caused by the helium rain which is followed by neon depletion (Wilson \& Militzer, 2010). 
The elements that are included in the calculation are hydrogen (solar), helium (solar), carbon (4.31 times solar), oxygen (value is allowed to vary, measured value by Galileo is 0.25 times solar), nitrogen (4.05 times solar), sulfur (2.88 times solar), argon (2.54 times solar), krypton (2.16 times solar), and xenon (2.11 times solar). The abundance of each component is taken from Atreya et al. (2003) based on the Galileo probe measurements. We then treat the water abundance as unknown and search for the $(O/H)_{J}$ required to reproduce $Z_{J}$ as derived by formation models. The results are listed in the last column of Table 1. From comparing the numbers it can be seen that that the upper bound estimate which assumes that all the heavies are represented by water does not differ by much from the one which uses the constraints introduced by the atmospheric measurements. 

An interesting thing to note is that in several cases oxygen is predicted to be sub-solar. In two extreme cases the total heavy element mass is so low that the predicted water abundance is found to be negative. 
These cases, however, are inconsistent with the measured abundances of the other elements and should not be taken as representative cases. Nevertheless, the cases with low water abundance emphasize the possibility of Jupiter being water-depleted. One can ask what is the minimum $Z_J/Z_{\odot}$ that is required to lead to positive water abundance based on the Galileo measurement. We find that $Z_J/Z_{\odot}$ must be equal or larger than 0.83 for $(O/H)_J/(O/H)_{\odot}$ to be positive. 
\par

Figure 2 shows the element ratios in Jupiter's envelope compared to solar as a function of Jupiter's global metallicity for various key components (eq.~2). Again, we use the Galileo measurements and each time assume that the abundance of a given component is unknown (while the rest are held fixed to their measured values). 
 We can then determine Jupiter's atmospheric (or envelope, assuming the planet is fully mixed) enrichment for a given species abundance.
Also here only hydrogen, helium, carbon, oxygen, nitrogen, sulfur, argon, krypton, and xenon are included. Since other elements are less abundant, their variation with increasing metallicity is less profound. In the figure shown are oxygen (solid), carbon (dashed), sulfur (dashed-dotted) and argon (dotted). While this is an oversimplified calculation that does not account for the relation between the various components, it shows clearly the sensitivity of the variation of the key elements with global metallicity. 

Since it is not clear that the water abundance in Jupiter can be set to the measured Galileo value due to its entry location but it is used in our computation when the variety of other elements is investigated, we consider two different abundances: 0.25 times solar, as determined by the Galileo probe (top panel) and 5 times solar (bottom panel). As can be seen from the figure, the oxygen (solid curve) and carbon (dashed curve) abundances are most sensitive to the planetary metallicity (due to their large fraction in solar abundance) while the variation in other species is relatively small. In addition, the expected value of the other components, for a given planetary metallicity, depends on the water abundance, a fact that strengthens the need to get an accurate water measurement. Finally, we suggest that in order to relate the heavy element enrichment in Jupiter with the abundances of individuals elements detailed chemical models which include the dependence of the various components with one another must be developed.

\subsection{Formation by Disk Instability}
An alternative model for giant planet formation is disk instability in which planets are formed from a gravitational instability in the disk (Boss, 1998). Recently, it has been shown that giant planets formed by this mechanism can also be enriched with heavy elements by various processes; enrichment from birth; mass-loss; and planetesimal capture (see Boley et al., 2011). So far, most of the studies have not concentrated on Jupiter but instead investigated the phenomenon globally. However, Helled and collaborators have addressed the case of Jupiter specifically. In Helled et al. (2006) the possible enrichment of Jupiter via planetesimal capture for a range of planetesimal sizes, velocities, and compositions, and solid-surface densities was investigated. The captured heavy element mass was found to be between 16.5 and 55.1 M$_{\oplus}$. This large range is due to the different cases that were assumed. While in Helled et al. (2006) Jupiter was assumed to form in situ, Helled \& Schubert (2009) explored how the captured mass changes with radial distance and disk properties. Again, due to the large number of free parameters, a large range was found: at 5 AU the captured heavy element mass was found to be 15-30 M$_{\oplus}$ for a disk mass of 0.05 M$_{\odot}$, and it was found to increase with increasing radial distance until $\sim$15 AU. 
The heavy element masses, and the derived $(O/H)_{J}$ from equations (1) and (2) for the disk instability model are also listed in Table 1. 
From comparing these values with the ones derived for core accretion it is suggested that in principle, the water measurement cannot be used to discriminate between the two formation models, at least not without additional constraints.  

\section{Interior Models}
The estimates for the heavy element mass in Jupiter are derived from interior models which use the measured physical properties of Jupiter such as its mass, its gravitational coefficients $J_{2n}$, the equatorial/mean radius, the 1 bar temperature, which is used to determine the planet's entropy, its rotation rate, and occasionally, the atmospheric He/H. A measurement of the global water abundance in Jupiter can be used as an additional constraint to interior models since if the {\it envelope metallicity} is set to its measured value, the parameter space of possible interior models decreases considerably (Guillot, 2005; Nettelmann et al., 2012). \par

While interior models are used to predict the range of heavy element mass in Jupiter, this range is relatively large due to the uncertainty in the equations of state of hydrogen and helium, their interaction, the composition of the heavy elements, and model parameters and assumptions such as the number of layers, the distribution of the heavy elements between the planets, the depth of differential rotation and the energy transport mechanism. Nevertheless, a range for the masses of heavy elements, which are typically represented by ices or rocks in interior models, can be derived (e.g., Fortney \& Nettelmann, 2010; Baraffe et al., 2014).  
A heavy element mass of 1-40 M$_{\oplus}$ was suggested by Saumon \& Guillot, (2004). Nettelmann et al. (2012) derived heavy element mass of $\sim$30 M$_{\oplus}$, both using three-layer models, but with different model assumptions and different equations of state. 
An alternative 2-layer interior model was presented by Militzer et al. (2008) who suggested that Jupiter's atmosphere is water-depleted with the water being concentrated in deeper regions, above the core, which was estimated to have a mass of 14-18M$_{\oplus}$. 
In this model the total mass of heavy elements was found to be 18 M$_{\oplus}$ with only 2 M$_{\oplus}$ mixed in Jupiter's envelope. To fit the measured $J_4$, differential rotation on cylinders had to be applied. 

Recently, Leconte \& Chabrier (2012) investigated Jupiter's internal structure assuming a non-adiabatic and in-homogenous structure. Inhomogeneous composition may occur due to mean molecular gradients that lead to a Õlayered convectiveÕ interior (Leconte \& Chabrier, 2012; 2013). With this configuration the internal temperatures are higher, and as a result, the total derived heavy-element mass is significantly larger than in adiabatic models. Jupiter's heavy element mass in the envelope was estimated to be 41 - 63.5 M$_{\oplus}$ with no (or very small) core. In that case, the amount of high-Z material, including water, in Jupiter's atmosphere does not represent the global enrichment of heavy elements throughout the planet. \par

We next estimate the $(O/H)_{J}$ based on interior model results. The methods and cases that are considered are similar to the ones described in the previous section. The results are summarized in Table 2. For the model of Leconte \& Chabrier, (2012) the derived $(O/H)_{J}$ are unrealistically large, but it should be noted that in this model the heavy elements are not homogeneously distributed, and therefore, the measurement of $(O/H)_{J}$ in a given region cannot be used to determine the global value. \par
\par

It should be noted that the possibility of  constraining the interior structure of Jupiter by Jovian seismology could be a promising independent method. While this idea is not new, only recently, oscillation patterns for Jupiter have been detected from the ground Gaulme et al. (2011). 
While the current data are not accurate enough to further constrain the interior structure of Jupiter, future improvements could do so and provide an independent measure of the density structure of the planet. Therefore, Jovian seismology seems to be a very promising method in giant planet studies and lead to a significant leap in our understanding of Jupiter.

\section{Water abundances below 30\%}
It is possible that the water abundance in Jupiter's atmosphere is actually relatively low. Indeed, several papers have argued that the measured low abundance of water in Jupiter's atmosphere is a signature of low water abundance in the planetesimals which seeded the planet. Lodders (2004) took the measured Galileo water abundance as the deep value and proposed that Jupiter has a carbon-to-oxygen ratio exceeding unity. It was argued that the point in the Solar Nebula at which water ice condensed--the snowline-- might have been further from the Sun than the point at which Jupiter formed, and therefore volatiles adhering to solid organics rather than water ice were carried into Jupiter. Thus in this model, the mass fraction of planetesimals that was water is zero, and the contribution to oxygen in Jupiter from water in planetesimals is formally zero. One challenge with this model is that the measured water abundance is sub solar, and yet Lodders (2004) brings in solar composition gas to account for noble gases. However, Stevenson and Lunine (1988) initially suggested that the nebula inward of the water snowline might be dried out by diffusive transport and the "cold-finger" effect of condensation. Since Lodders suggests that Jupiter could have been inward of the snowline, the inference that the nebular gas was depleted of water is consistent both with her interpretation of a subsolar oxygen abundance in the Jovian envelope and the nebular cold-finger effect. 

A somewhat less extreme model is that of Mousis et al. (2012), who fitted the Galileo probe data by modeling planetesimals forming in an oxygen-- depleted disk (C/O of unity). Because the oxygen not in rocks is bound primarily to CO rather than to hydrogen, and there is very little CO$_2$, the resulting mass fraction of water in the planetesimals is again essentially zero as in Lodders (2004). However, a significant amount of oxygen is brought in as CO, equal essentially to the carbon abundance. 
The scenario of low water abundance in Jupiter is indicated in the dashed-dotted curve in the bottom panel of figure 1 in which water is assumed to consist of only 15\% of the total heavy element mass. Since in Jupiter's envelope the CO and CO$_2$ will largely be converted to H$_2$O, we can estimate the water abundance if the total heavy element mass is inferred. For example, we predict that for a water-depleted Jupiter the O/H will be about 1.5 times solar for M$_z \sim$ 20 M$_{\oplus}$, and twice solar for M$_z$ = 28 Earth masses.

\section{The Challenges}
There are several challenges associated with {\it Juno}'s water abundance measurement and its interpretation: 
\begin{itemize}
\item The measurement itself is limited to the 100-200 bar pressure-level, and it is not clear whether it indeed represents the {\it global} water abundance in Jupiter. In order to estimate the global water abundance one has to {\it assume} that the heavy elements are homogeneously mixed. Although adiabatic evolution of Jupiter is found to be consistent with its current radius and luminosity, the possibility of non-adiabatic internal structure remains. 
\item An additional complication in the Juno water measurement is our uncertain knowledge of the microwave line shapes of ammonia and water as well as the deep abundance of ammonia. Indeed, it was argued by de Pater et al. (2005) that an orbiting microwave radiometer could only constrain water with large uncertainties. Janssen et al. (2005) demonstrated that, given the knowledge of the opacities at that time, the microwave radiometer (MWR) on Juno would readily distinguish both large enrichments of water relative to solar as well as gross under-abundances, but would have difficulty distinguishing among values closer to solar--for example, between solar and twice solar. More recently, Karpowicz \& Steffes (2011) have completed a more detailed determination of the water opacity in the relevant wavelength range, and Duong et al. (2014) have done likewise for the opacity contributed by ammonia-water clouds.  Although the implications for the Juno measurements of water abundance have yet to be quantified, we expect the result to be improved accuracy for values around the solar abundance.
\item Although a determination of the water abundance in Jupiter can be linked to the planetesimals' properties, the problem is rather degenerate.  For example, how can we distinguish small planetesimal sizes from low material strength and moderate velocities? All three cases result in faster and more efficient accretion of icy planetesimals. It's therefore important to investigate whether there's a way to quantify the different enrichments based on different assumed planetesimal properties.  
\item A missing piece in planet formation models is the location within the planet in which the accreted planetesimals are deposited, and how this distribution changes with time. If the heavy elements tend to reach the center, the envelope enrichment will be significantly smaller than in the case in which most of the accreted high-Z material is mixed within the envelope. A detailed investigation of this topic could be used to better connect formation models with the current internal structure of Jupiter.  
\item Can the $(O/H)_{J}$ measurement constrain Jupiter's formation location? The suggestion that Jupiter's enrichment with noble gases indicates formation farther out, where disk temperatures are significantly lower (Owen et al., 1999) cannot be confirmed easily since there is no easy way to distinguish formation at large radial distances and the delivery of planetesimals from outer regions.  The model of Lodders (2004), on the other hand assumes little acquisition of planetesimals from large radial distances.
\item Predictions from both formation and interior models suggest that it is possible that the mass of heavy elements in Jupiter can be about or less than $\sim$ 5 M$_{\oplus}$. 
If this is the case, the $(O/H)_{J}$ can be solar, or even sub-solar. With {\it Juno}'s  Microwave Radiometer it will be challenging to distinguish between different values of enrichment if the atmosphere is water-depleted. While it is not possible to determine whether this scenario is relevant for Jupiter, it might still be valuable to develop methods to distinguish among different enrichments of water in water-poor atmospheres. 
\item Can Jupiter be water-poor but oxygen-rich? It is possible for the protoplanetary disk to have been water poor in places and at times, while Jupiter accreted oxygen not only in rocks but in molecules such as CO and CO$_2$. However, the expectation  that re-equilibration of the molecular species in the Jovian envelope be fast is a reasonable one, unless Jupiter were extremely stratified.
\end{itemize}

\section{Implications for Saturn and extrasolar planetary systems}

A successful measurement of water by {\it Juno} at Jupiter will make a compelling case for a similar measurement at Saturn. In the case of a low water abundance--subsolar to about twice solar--the question will arise as to whether Saturn has a similarly low value. Abundances in Saturn come from remote sensing as no entry probe mission has been conducted to the ringed world, and its colder temperatures result in condensation of species at correspondingly deeper levels than in Jupiter. The most reliable heavy element measurement is that of carbon, which is 10 to 11 times solar (Fouchet et al, 2009)--thus triple the Jovian value. If oxygen turns out to be triple the Jovian value as well, then a tight coupling between the two would imply trapping of the carbon within molecular carriers such as CO--that is, the model of Mousis et al. (2012) . If, on the other hand, oxygen does not increase as much from the Jovian value, or does not increase at all, some process decoupling water from the carbon phases may be indicated--such as the carbon-rich planetesimals of Lodders  (2004). Alternatively, if Saturn did not form in a water-depleted region whereas Jupiter did, one might expect the Saturnian water enrichment to exceed that of the carbon--that is, be greater than 10. For example, if Lodders (2004) is correct that Jupiter formed inwards of the nebular snowline, it does not imply the same for Saturn--indeed, Saturn plausibly might have formed beyond the snowline, from planetesimals with abundant water, resulting in a relatively larger increase in the oxygen enrichment (relative to carbon) from Jupiter to Saturn. While all of these comparative considerations provide compelling reason for sampling water in the deep atmosphere of Saturn, this will be much more difficult than for Jupiter. A microwave radiometer will need to penetrate to hundreds of bars, through substantially larger opacity sources than for Jupiter, in order to measure water.

The increasing success in obtaining spectral signatures of oxygen- and carbon-bearing species in the atmospheres of extrasolar planets (Deming et al., 2013) bodes well for eventual comparison of the ratios of these from planetary system to system. Variations in elemental C/O ratios between stars and their planetary companions in principle reveal details of formation processes in the original planet-forming disks (Johnson et al. 2012), but to understand the connection will require systematic surveys as a function of stellar metallicity, planet mass, and number of planets. Giant planets within the same planetary system may exhibit different C/O values according to where they formed relative to condensation fronts in their natal disks (Ali-Dib et al., 2014), perhaps providing a record of formation location that would otherwise have been obscured by subsequent migration. This makes spectroscopy of hot Jupiters by Hubble Space Telescope and, in the near future, James Webb Space Telescope, of keen interest. 
 
\section{Discussion and Conclusions}
The measurement of Jupiter's water abundance from microwave spectroscopy up to pressures of $\sim$ 100 bars by {\it Juno} will provide valuable constraints on formation and structure models of Jupiter. 
A complication arises, however, due to the uncertainty in the distribution of the heavy elements in Jupiter, and as a result, whether this measurement represents the global enrichment of the planet. In addition, there is no simple way to relate the water abundance in Jupiter and its origin. Even within the assumption that Jupiter had formed via core accretion, different model assumptions can result in very different final compositions, and given the large number of free parameters and unknowns, it is not straightforward how to connect the water abundance to the planetesimals' properties and the location in which Jupiter had formed and/or to the physical/chemical conditions of the solar nebula. \par 

Once the water abundance is measured by {\it Juno}, models are expected to face a true challenge, unless Jupiter is found to have $(O/H)_J$ significantly higher than solar. This is because current formation and structure models suggest a large water enrichment, of the order of ten times solar which seems to be too large. What if {\it Juno} measures a water enrichment which is similar to the other components as derived by the Galileo probe? In that case, $(O/H)_{J}$ will be about 2-3 times solar and by looking at the tables, an immediate conclusion will be that the total heavy element mass in Jupiter must be significantly lower than the one predicted by {\it both} formation and interior models. If the water enrichment is found to be solar, i.e., smaller compared to other materials, it may imply that the water (i.e., oxygen) in Jupiter is hidden closer to the center (Militzer et al., 2008), however, in that case an explanation for the different behavior of oxygen vs. carbon, silicon, and other components, which is not trivial,  will be required. Alternatively, a low water enrichment may reflect low abundances of water in the solar nebula --either the result of elevated $(C/O)$ or drying effects along the nebular snowline. 

It is clear that in order to make full usage of {\it Juno}'s measurements several investigations are required. It is desirable to define the relations between various model assumptions and the derived composition, and to investigate the different inferred compositions for various formation scenarios: for example, in situ formation vs. migration, core accretion vs. disk instability, and different planetesimal properties. Finally, there is a clear need to derive interior models of Jupiter in which a large parameter space of possibilities are considered (equations of state, heat transport mechanisms, rotational state, number of layers, in/homogeneity in composition). Progress in these directions will be useful for interpretation of {\it Juno} data as well as for understanding gas giant planets in general. 

\section*{Acknowledgments}
We thank Mark Marley for valuable comments. We also thank the members of the {\it Juno} science team and Peter Bodenheimer, Attay Kovez, Akiva Bar-Nun and Sushil Atreya for fruitful discussions and helpful suggestions. This work was supported by the {\it Juno} Project.

\clearpage
\begin{figure}
\begin{center}
\includegraphics[angle=0,height=9cm]{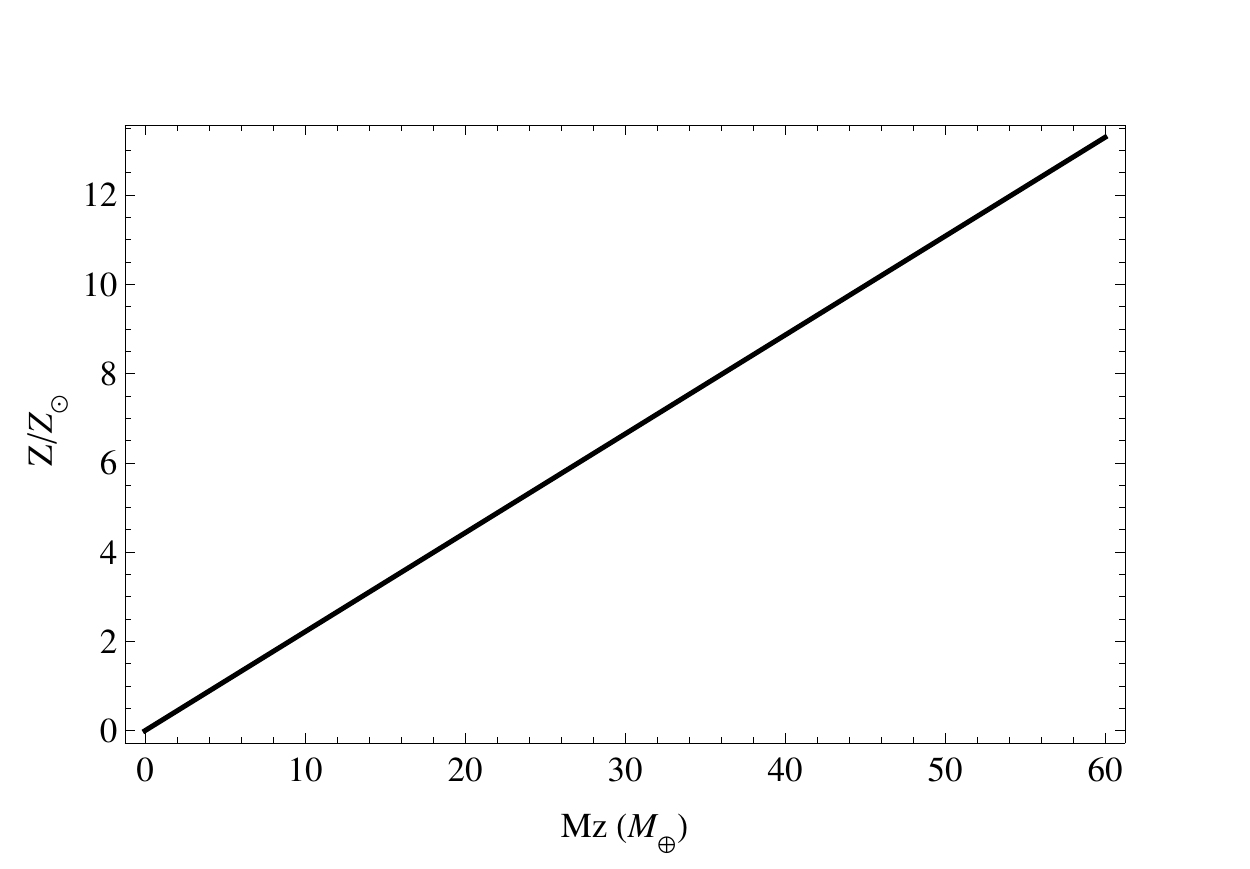}
\includegraphics[angle=0,height=9cm]{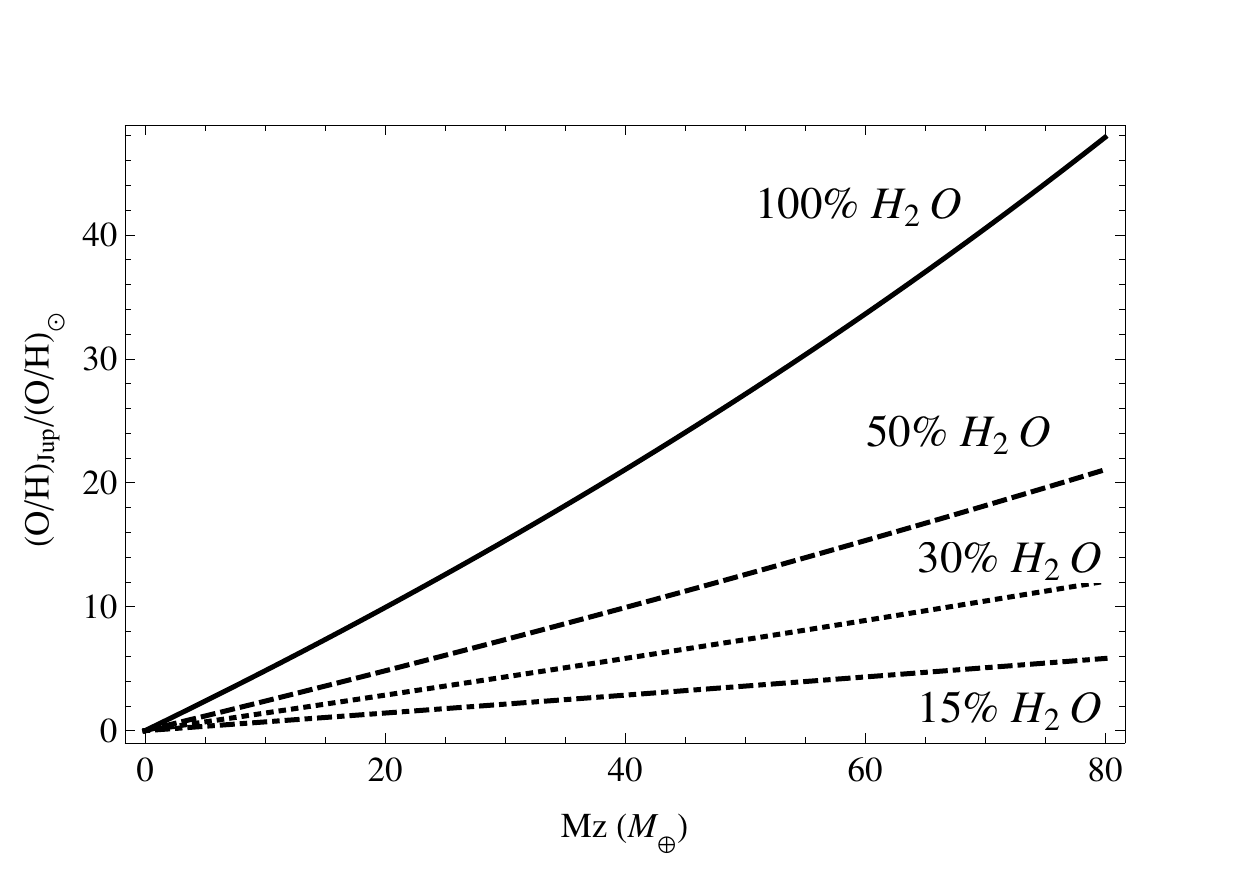}
\caption{{\bf Top:} Jupiter's metallicity divided by stellar metallicity ($Z/Z_{\odot}$) vs.~heavy-element mass ($M_z$) in Earth masses (M$_{\oplus}$). {\bf Bottom:} $\frac{(O/H)_{J}}{(O/H)_{\odot}}$ vs.~heavy-element mass ($M_z$) in Earth masses for planetesimals composed of 100\% (solid curve) 50\% (dashed curve), 30\% (dotted curve) and 15\% (dashed-dotted curve) water by mass using  Eq. 1.}
\label{fig:genn}
\end{center}
\end{figure}

\clearpage
\begin{figure}
\begin{center}
\includegraphics[angle=0,height=18cm]{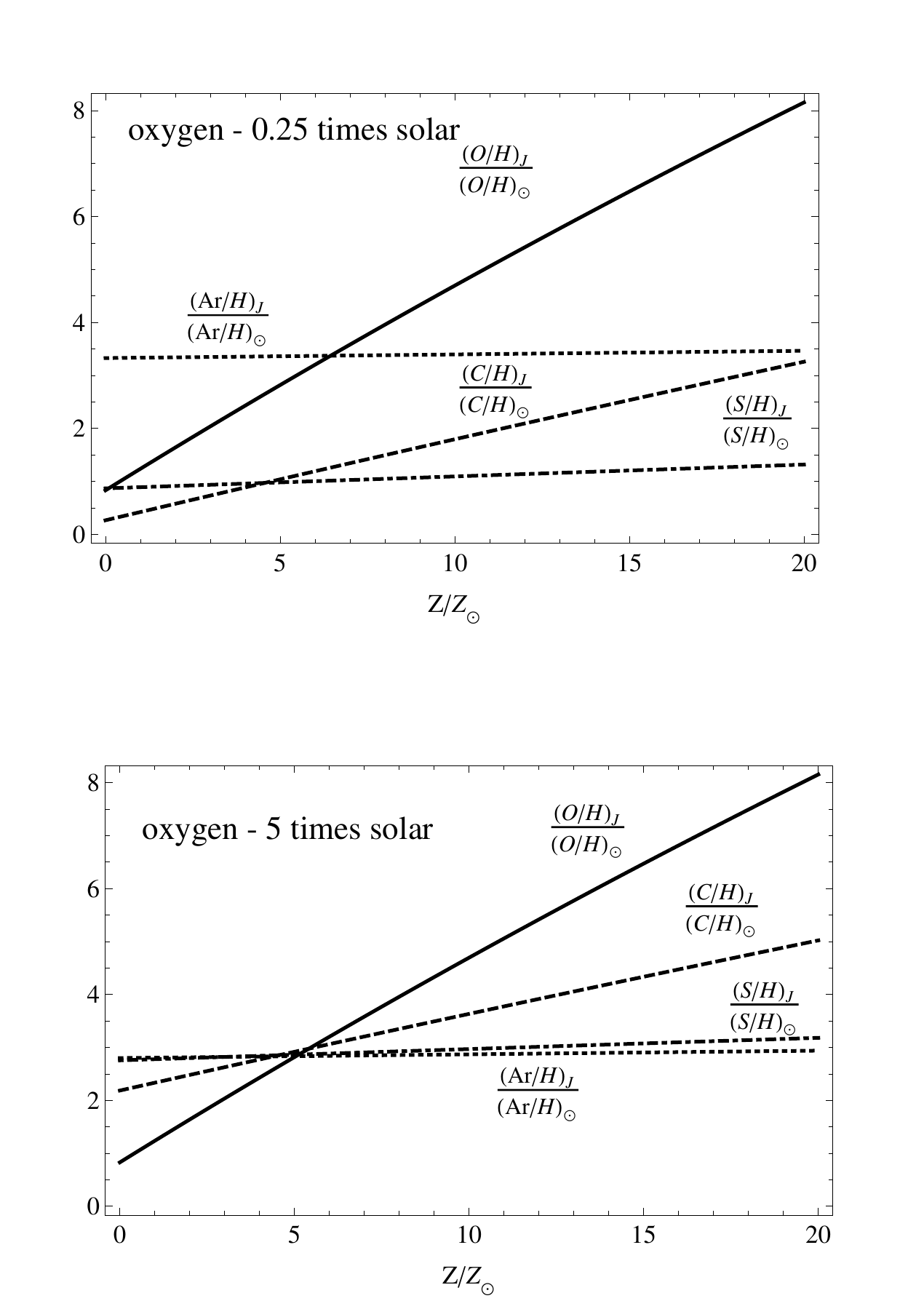}
\caption{Element ratios in Jupiter's atmospheres compared to solar as a function of Jupiter's (total) metallicity for various key elements when the oxygen abundance is set to 0.25 times solar ({\bf Top}) and 5 times solar ({\bf Bottom}). Each time we assume that the abundance of a given component is unknown (while the rest are held fixed to the value measured by Galileo) and compute the variation of a given element to the planetary metallicity. Shown are the results for oxygen (solid), carbon (dashed), sulfur (dashed-dotted) and argon (dotted).}
\label{fig:genn}
\end{center}
\end{figure}

\clearpage
\begin{sidewaystable}[h!]
\footnotesize
 \begin{tabular}{|c|c|c|c|c|c|c|c|}
 \hline
 Reference & Heavy Element Mass (M$_{\oplus}$) & $Z_{J}/Z_{\odot}$ & MAX $\frac{(O/H)_{J}}{(O/H)_{\odot}}$, eq. (1) & $\frac{(O/H)_{J}}{(O/H)_{\odot}}$, eq. (1) 30\% $H_2O$  & $\frac{(O/H)_{J}}{(O/H)_{\odot}}$, eq. (2)\\
\hline
\hline
Alibert et al. (2005b)$^{(1)}$   & 28 & 6.2 & 14.2 & 4.78& 14.2  \\
\hline
Alibert et al. (2005b)$^{(2)}$   & 32 & 7.1 & 16.4 & 5.54 & 16.8  \\
\hline
Lissauer et al. (2009)$^{(1)}$  &  5 & 1.1 & 2.4 & 0.79 & 0.67 \\
\hline
Lissauer et al. (2009)$^{(2)}$  &  15 & 3.3 & 7.3 & 2.46 & 6.3 \\
\hline
Helled et al. (2006) & 16.5 - 55.1 & 3.6 - 12.2 & 8.1 - 30.4 & 2.71 - 10.33 & 7.1 - 33.2  \\
\hline
Helled \& Schubert (2009) & 15 - 30 & 3.3 - 6.6 & 7.3 - 15.3 & 2.46 - 5.15 &6.3 - 15.4 \\
\hline
 \end{tabular}
 \caption{Formation Models: Heavy element mass in Jupiter's envelope, envelope metallicity, and derived $\frac{(O/H)_{J}}{(O/H)_{\odot}}$. Case (1) corresponds to the heavy elements in the envelope assuming that the during phase 1 all the planetesimals reach the core, while case (2) assumes that once the core mass reaches 2 M$_{\oplus}$ the rest of the accreted heavy elements are dissolved in the envelope, therefore, resulting in larger envelope enrichments. Note that the heavy element mass and the corresponding metallicity $Z_{J}/Z_{\odot}$ correspond only to the envelope, and do not include the core.  Solar metallicity $Z_{\odot}$ is set to be 0.0142.  The estimates from Helled \& Schubert (2009) correspond to formation at 5.2 AU, in a disk with mass of 0.05M$_{\odot}$ and three different density profiles.}
\end{sidewaystable}

\begin{sidewaystable}[h!]
\footnotesize
 \begin{tabular}{|c|c|c|c|c|c|c|c|}
\hline
Reference & Heavy Element Mass (M$_{\oplus}$) & $Z_{J}/Z_{\odot}$ & $\frac{(O/H)_{J}}{(O/H)_{\odot}}$, eq. (1)  & $\frac{(O/H)_{J}}{(O/H)_{\odot}}$, eq. (1) 30\% $H_2O$ & $\frac{(O/H)_{J}}{(O/H)_{\odot}}$, eq. (2)\\
\hline
\hline
Saumon \& Guillot (2004) & 1 - 40 & 0.2 - 8.9 & 0.47 -  21.1 & 0.16 - 7.1 & negative - 20.2 \\
\hline
Nettelmann et al. (2012) &  28 - 32  & 6.2 - 7.1 & 14.2 - 16.4 & 4.8 - 5.54 & 14.2 - 16.8  \\
\hline
Militzer et al. (2008) & 0-4 & 0 - 0.9 & 0 - 1.9 & 0 - 0.63& negative - 0.2  \\
\hline
Leconte \& Chabrier (2012) & 41 - 63.5 & 9.1 - 14.1  & 21.6 - 36.0 & 7.3 - 12.3  & 22.9 - 40 \\
\hline
 \end{tabular}
 \caption{Same as Table 1 but for structure models. The values listed in the table correspond to a case in which all the heavy elements in Jupiter within the envelope are assumed to be homogeneously mixed. 
A measurement of $(O/H)_J$ which is lower than the listed values could indicate an inhomogeneity in the high-Z material distribution within Jupiter's interior (see e.g., Nettelmann et al., 2012).}
\end{sidewaystable}
\clearpage

\end{document}